\documentclass[pra,twocolumn,superscriptaddress,showpacs,10pt]{revtex4}
\usepackage{graphicx}
\usepackage{caption2}

\begin{document}

\title{Nonadiabatic production of spinor condensates with a QUIC trap}
\author{P. Zhang}
\affiliation{School of Physics, Georgia Institute of Technology, Atlanta, Georgia 30332,
USA}
\author{Z. Xu}
\affiliation{Center for Advanced Study, Tsinghua University, Beijing 100084, People's
Republic of China}
\author{L. You}
\affiliation{School of Physics, Georgia Institute of Technology, Atlanta, Georgia 30332,
USA}
\affiliation{Center for Advanced Study, Tsinghua University, Beijing 100084, People's
Republic of China}
\date{\today}

\begin{abstract}
Motivated by the recent experimental observation of
multi-component spinor condensates via a time-dependent
quadrupole-Ioffe-configuration trap (QUIC trap), we provide a
general framework for the investigation of nonadiabatic
Landau-Zener dynamics of a hyperfine spin, e.g., from an atomic
magnetic dipole moment coupled to a weak time-dependent magnetic
(B-) field. The spin flipped population distribution, or the
so-called Majorona formula is expressed in terms of system
parameters and experimental observables; thus, provides much
needed insight into the underlying mechanism for the production of
spinor condensates due to nonadiabatic level crossings.
\end{abstract}

\pacs{32.80.Bx, 03.75.Lm, 32.60.+i}
\maketitle

\section{Introduction}

Magnetic traps play an important role in the study of atomic
Bose-Einstein condensates (BEC) \cite{magnetictrap}. In a typical
static magnetic trap, individual atomic spin couples to the
spatial dependent magnetic (B-) field because of Zeeman effect.
When an atom moves in a region where the direction of the B-field
changes slowly and the strength of the B-field is sufficiently
strong, according to Born-Oppenheimer approximation
\cite{bo1,bo2}, the atomic spin can follow the B-field
adiabatically and remain in the same trapped eigen-state of the
interaction Hamiltonian relative to the
instantaneous direction of the magnetic field ${\vec B}({\vec r})$, where ${%
\vec r}$ is the center of mass position of the atom (or more
precisely, that of the valence electron). In this case the atomic
center of the mass experiences an effective spatially-varying
potential that is equal to the Zeeman energy and proportional to
the strength of the B-field.

For weak B-fields, when the atomic Zeeman energy is comparable to
or less than the frequency of the directional variation of the
B-field felt by the moving atom, adiabatic dynamics cannot be
followed anymore. As a result, nonadiabatic (Majorona) transitions
\cite{Majorona} for the atomic spin may occur. Two potentially
damaging effects can cause nonadiabatic transitions. The first
happens when an atom enters a weak B-field region due to its
translational motion in space. For instance, in a quadrupole trap,
atoms in the weak field seeking state are accelerated towards the
center of the trap where the B-field vanishes. Nonadiabatic
transitions always occur in the vicinity of a zero B-field. To
avoid this region of vanishing B-field or a spatial "hole," a
number of methods have been developed to effectively plug it,
e.g., with the use of a far-off-resonant optical potential as an
``optical plug" \cite{opticalplug} or the more famous time
averaged orbiting potential (TOP) trap \cite{top}. The second
reason for nonadiabatic transitions is due to the explicit time
dependence of the B-field. Obviously nonadiabatic transitions may
occur if the B-field changes rapidly with time.

Recently, the atomic quantum gas group at Peking University (PKU)
reported interesting observations of multi-component $^{87}$Rb
($F=2$) spinor condensates via switching off the B-fields of an
initially spin polarized single component condensate in a QUIC
trap \cite{chenshuai}. The group of Prof. Chandra Raman at Georgia
Tech also discovered counter-intuitive meta-stability when
condensates were loaded into an ``unplugged" magnetic quadruple
trap \cite{raman}. We decided to present our theoretical studies
in the hope that the theoretical framework for spinor nonadiabatic
level crossing dynamics may be of interest to other groups in the
field of atomic quantum gases. In this paper, we will focus on the
Peking University experiment in a time-dependent QUIC trap
\cite{hansch}. The more involved situation of a condensate in a
quadruple trap will be discussed elsewhere \cite{peng}. According
to the reported experiment \cite{chenshuai}, the affected
time-dependence for the B-field is relatively simple. After a
single component condensate was created in a QUIC trap, the
various B-field generating currents were switched off in
appropriately chosen orders. Whenever near-zero level-crossing
occurs, multi-component spinor condensates are observed.

This paper summarizes our treatment of level crossing dynamics for
an atomic spin inside an external B-field. The theory is developed
with respect to ``the first scenario," where the vanishing B-field
is due to the different time constants of decay for the B-fields
from the QUIC coil and the bias coil after being shut off as
discussed in Sec. II. An alternative scenario where the B-field
zero is due to different time constants of the decaying B-fields
from the quadruple coil and the Ioffe coil will be discussed in
Sec. III. Finally we conclude and provide a brief summary in Sec.
IV.

\begin{figure}[h]
\includegraphics[width=5.2cm,height=4.5cm]{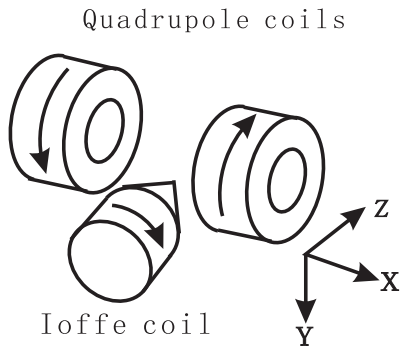} \vspace{0.5cm}
\caption{ The QUIC trap geometry (excluding the bias coils). The arrows
indicate directions of currents in the coils. }
\label{fig1}
\end{figure}

\section{The First Scenario}

The magnetic trap used in their experiment \cite{chenshuai} is
made up from two separate coils, a QUIC coil and a bias coil. The
QUIC coil consists of a quadruple trap coil and an Ioffe coil in
series as in the original QUIC trap \cite{hansch}. The
compensating coils for the earth's B-field are separate and always
left on; thus will not be included explicitly in our model. Before
switching off, the magnetic B-field ${\vec{B}}^{\rm Q}({\vec{r}})$
created by the QUIC coil has the familiar configuration of a
Ioffe-Pritchard trap and can be expressed as
\cite{chenshuai,hansch}
\begin{eqnarray}
{\vec{B}}^{\rm Q}({\vec{r}})=B^{\rm Q}_{\perp }({%
\vec{r}})\mathbf{e}_{\mathbf{\perp }}+B^{\rm Q}_{z}({%
\vec{r}})\mathbf{e}_{\mathbf{z}},
\end{eqnarray}
where the axial and radial QUIC B-field components are  $B^{\rm Q}_{\perp }({%
\vec{r}})$ and $B^{\rm Q}_{z }({%
\vec{r}})$, respectively;
\begin{eqnarray}
B^{\rm Q}_{z}({\vec{r}})
&=&B^{\rm Q}_{z}(0)+B^{\rm Q\prime\prime}_{z }z^{2},\nonumber\\
B^{\rm Q}_{\perp }({\vec{r}}) &=&B^{\mathrm{Q\prime}%
}_{ \perp  }\sqrt{x^{2}+y^{2}}, \label{ioffe}
\end{eqnarray}%
with the unit vector $\mathbf{e}_{\mathbf{\perp }}$ defined as
\begin{eqnarray}
\mathbf{e}_{\mathbf{\perp }}=\frac{(-x\mathbf{e}_{x}+y\mathbf{e}_{y})}{%
\sqrt{x^{2}+y^{2}}}.
\end{eqnarray}
$B^{\mathrm{Q\prime}}_{\perp}$ and $B^{\rm Q\prime\prime}_{z}$
denote the respective spatial derivatives for the B-fields. The
right handed coordinate system as in Figure \ref{fig1} is chosen
such that $B^{\rm Q}_{z}(0)>0$ and $B^{\mathrm{Q\prime}}_{ \perp
}>0$.

The bias field ${\vec{B}}^{\rm A}$ is in the $z$ direction. It is
created by the bias coils and approximately constant near the trap
center \cite{chenshuai}.
Before switching off, it can be expressed as ${\vec{B}}^{\rm A}({\vec{r}}%
)=-B^{\rm A}_{ z}\mathbf{e}_{\mathbf{z}}$ satisfying $B^{%
\rm Q}_{z }(0)>B^{\rm A}_{ z}>0$. If the bias field is switched
off first and the two components of the QUIC
field are simultaneously switched off after a time interval of $t_{\mathrm{%
int}}$, then at time $t$, the QUIC field becomes $e^{-t/\tau_{\rm Q}}{\vec{B}}^{%
\rm Q}({\vec{r}})$ and the bias field becomes
$e^{-(t+t_{\mathrm{int}})/\tau_{\rm A}}{%
\vec{B}}^{\rm A}({\vec r})$, i.e., both the QUIC field and the
bias field are assumed to decrease exponentially with decay time
constants $\tau_{Q}$ and $\tau_{A}$, and the quadruple field and
the Ioffe field are assumed to decay with the same time constant.
Assuming $t=0$ as the instant for shutting off the QUIC field, the
total time-dependent B-field then becomes
\begin{eqnarray}
{\vec{B}}({\vec{r}},t) &=&e^{-t/\tau_{Q}}B^{\rm Q}_{\perp }({\vec{%
r}})\mathbf{e}_{\mathbf{\perp }}  \nonumber \\
&&+e^{-t/\tau_{Q}}B^{\rm Q}_{z}({\vec{r}})\mathbf{e}_{\mathbf{z}%
}-e^{-(t+t_{\mathrm{int}})/\tau_{\rm A}}B^{\rm A}_{ z}\mathbf{e}_{\mathbf{z%
}}.
\end{eqnarray}%
$B^{\rm Q}_{\perp }({\vec{r}})$ is proportional to $%
(x^{2}+y^{2})^{1/2}$ near the trap center or the origin; therefore, we have $B^{%
\rm Q}_{z }({\vec{r}}),\ B^{\rm A}_{z}\gg B^{\rm Q}_{\perp }({\vec{r}})$ and $|B^{\rm Q%
}_{z}({\vec{r}})-B^{\rm A}_{z}|\gg B^{\rm Q}_{\perp }({%
\vec{r}})$. Before switching off the QUIC field, the $z$ component of ${\vec{%
B}}({\vec{r}},t)$ takes a positive value $B^{\rm Q%
}_{ z }({\vec{r}})-B^{\rm A}_{ z}$ much larger than the initial
value of the transverse field $B^{\rm Q}_{\perp}({\vec{r}})$. At
$t=0$, all atomic spins initially are polarized, thus are the
eigen-state $|M_{F}=2\rangle $ of $F_{z},$ i.e., the $z$ component
of the atomic hyperfine spin $\vec{F}$. If the QUIC field and the
bias field are switched off simultaneously, i.e.,
$t_{\mathrm{int}}=0$ and $\tau_{Q}=\tau_{A}$, the direction of the
total B-field ${\vec{B}}({\vec{r}},t)$ does not change with time
although the strength of ${\vec{B}}({\vec{r}},t)$ decreases after
the switching off process. Nonadiabatic transitions do not occur
in this case and the initial single component condensate remains a
single component one. If the QUIC field and the bias field
decrease with different time constants $\tau_{\rm Q}\neq \tau_{%
\rm A}$, the direction of ${\vec{B}}({\vec{r}},t)$ changes with
time and nonadiabatic level crossing arises.

In the calculations to follow, we will make a simple approximation
that the atomic spatial position does not change during the
switching-off process. This allows for an easy calculation of
nonadiabatic transition probabilities between different atomic
spin states at a fixed spatial position ${\vec r}$. This is well
justified for the experiment of PKU, where level crossing occurs
over a time window of $\sim 10^2(\mu {\mathrm s})$, during which
an condensed atom moves a distance less than $0.1(\mu {\mathrm
m})$, provided its kinetic energy is $\sim 10^{4}({\mathrm
{Hz}})$.

As mentioned above, the $z$ component of the B-field initially
takes a large positive value. After the switching-off, the bias
field decreases much slower than the QUIC field \cite{chenshuai},
i.e. we have $\tau_{\rm A}\gg \tau_{\rm Q}$. At certain instant $%
t_{0}$, the value of $e^{-t/\tau_{Q}}B^{\rm Q}_{z}$ equals
$e^{-(t+t_{\mathrm{int}})/\tau_{\rm A}}B^{\rm A}_{ z}$, which
causes the $z$ component of the total B-field to become zero. As a
result of this, transitions from the state $|M_{F}=2\rangle $ to
other eigen-states of $F_{z}$ occur because of the finite
transverse B-field $e^{-t/\tau_{Q}}B^{\rm Q}_{\perp }$ in the
vicinity of $t_{0}$. After $t_{0}$, the $z$ component of the
B-field becomes negative because
$e^{-(t+t_{\mathrm{int}})/\tau_{\rm A}}/e^{-t/\tau_{Q}}\gg 1$ for
a large enough $t$; the absolute value of the $z$ component of the
B-field can  become again much larger than the transverse
components of ${\vec{B}}({\vec{r}},t)$ for $t\gg t_{0}$.
Therefore, the probabilities for an atom in different eigen-states
of $F_{z}$ can again take constant values in the long time limit.

To compute the nonadiabatic level crossing rates, we note that
transitions mainly occur in the near zero B-field region, i.e.,
for weak B-field. Thus, we only need to consider the linear Zeeman
coupling of an atomic hyperfine spin. Our model Hamiltonian takes
the simple form
\begin{eqnarray}
H=g_{F}\mu_{{\mathrm B}}{\vec B}({\vec r},t)\cdot \vec F.
\label{Hamiltonian}
\end{eqnarray}
Here $g_{F}$ is the {\it Land$\acute{\rm e}$} $g$ factor and
$\mathrm{\mu }_{\mathrm{B}}$ is the Bohr magneton. For $^{87}$Rb
atoms under consideration here, the spinor degree of freedom
refers to the $F=2$ manifold with $g_F=1/2$. In their experiment \cite%
{chenshuai}, the initial condition corresponds to
\begin{eqnarray}
|\Psi (0)\rangle =|M_F=2\rangle.
\end{eqnarray}
At large $t\rightarrow+\infty$, the wave function can be expanded
as
\begin{eqnarray}
|\Psi(t\rightarrow+\infty)\rangle =\sum_{M_F=-2}^{2}C_{M_F}( {\vec
r}) e^{-i\phi_{M_F}({\vec r},t)}|M_F\rangle,
\end{eqnarray}
in the complete basis of $F_z$ along the initial quantization
axis. Our problem is to find the steady population distribution
$P_{M_F}( {\vec r}) =|C_{M_F}({\vec r})|^{2}$ in the long time limit.

We will make use of the method of Hioe \cite{hioe} to
calculate the finial state probability distribution due to
nonadiabatic level crossing of a high spin. Because of
the rotational symmetry of our model system (\ref{Hamiltonian}), it can be
mapped onto a spin $1/2$ spinor with the same type of coupling, described by
a Hamiltonian
\begin{eqnarray}
H_{\sigma}=g_F\mu_{\mathrm B}{\vec B}(\vec r,t)\cdot
{\frac{\vec\sigma}{2}}, \label{H12}
\end{eqnarray}
where $\vec\sigma$ is the familiar spin $1/2$ Pauli matrix vector. The
initial condition for the spin $1/2$ state is
\begin{eqnarray}
|\varphi (0)\rangle =[1,0]^{T},
\end{eqnarray}
and the finial state can be denoted as
\begin{eqnarray}
|\varphi(t\rightarrow+\infty)\rangle =[\alpha ({\vec
r})e^{i\phi_{\alpha}({\vec r},t)},\beta ({\vec
r})e^{i\phi_{\beta}({\vec r},t)}]^{T}.
\end{eqnarray}
Upon solving this two state problem, $P_{M_F}({\vec r})$ can be
found easily  according to the rotation group representation
elements as in Hioe \cite{hioe}. Apart from a globe phase factor,
the evolution operator corresponding to the Hamiltonian
(\ref{Hamiltonian}) can be expressed as
$D^{(2)}=\exp[-i\hat{n}\cdot\vec{F}\theta]$; while the one
corresponding to the Hamiltonian (\ref{H12}) is
$D^{(1/2)}=\exp[-i\hat{n}\cdot(\vec{\sigma}/2)\theta]$. The unit
vector $\hat{n}$ and the angle $\theta$ are determined by
$\vec{B}(\vec{r},t)$. Therefore, $D^{(1/2)}$ and $D^{(2)}$ are the
representation matrixes (D matrixes) of the {\it same} rotation
operation. The transition probabilities $P_{M_F}$ and
$|\alpha(\vec{r})|^{2}$ can be rewritten as
$P_{M_F}=|D^{(2)}_{M_F,2}|^{2}$ and
$|\alpha(\vec{r})|^{2}=|D^{(1/2)}_{1/2,1/2}|^{2}$. According to
the representation theory of ${\rm SO}(3)$ group \cite{winger},
$|D^{(2)}_{1/2,1/2}|$ and $|D^{(2)}_{M_F,2}|$ are functions of
$\sin[\beta/2]$ and $\cos[\beta/2]$ with $\beta$ one of the three
Euler angles of the rotation. Although we do not know the values
of $\hat{n},\theta$, and $\beta$, we can express the transition
probability $P_{M_F}$ in terms of $A({\vec r})=|\alpha ({\vec
r})|^{2}$; as,
\begin{eqnarray}
P_{2}({\vec r}) &=&A({\vec r})^{4},  \nonumber \\
P_{1}({\vec r}) &=&4A({\vec r})^{3}[1-A({\vec r})],  \nonumber \\
P_{0}({\vec r}) &=&6A({\vec r})^{2}[1-A({\vec r})]^{2},  \label{probability}
\\
P_{-1}({\vec r}) &=&4A({\vec r})[1-A({\vec r})]^{3},  \nonumber \\
P_{-2}({\vec r}) &=&[1-A({\vec r})]^{4}.  \nonumber
\end{eqnarray}

The two state problem can be solved accurately with the
Landau-Zener formula \cite{lz}. To this end, we reexpress the
Hamiltonian (\ref{H12}) as
\begin{eqnarray}
H_{\sigma }[t]=q[t]h_{\sigma}[t],
\end{eqnarray}
with the "normalized" Hamiltonian
\begin{eqnarray}
h_{\sigma}[t]=g_{\perp }\sigma _{\perp }+\left( g^{\rm Q}_{ z }-%
e^{-\xi t}g^{\rm A}_{ z }\right) \sigma _{z},
\end{eqnarray}
and the parameters
\begin{eqnarray}
g_{\perp } &=&\frac{1}{2}g_{F}\mu_{\mathrm B}B^{\rm Q}_{\perp }({%
\vec{r}}), \nonumber \\
g^{\rm Q}_{ z } &=&\frac{1}{2}g_{F}\mu_{\mathrm
B}B^{\rm Q}_{z}({\vec{r}}),  \nonumber \\
g^{\rm A}_{ z } &=&\frac{1}{2}g_{F}\mu_{\mathrm B}e^{-t_{{\mathrm {int}}}/\tau_{{\mathrm A}}}B^{%
\rm A}_{z }({\vec{r}}), \nonumber \\
\xi &=&(\tau_{\rm A}-\tau_{\rm Q})/{\tau }{_{\rm A}}%
{\tau }{_{\rm Q}}. \label{g}
\end{eqnarray}%
We define a new time variable
\begin{eqnarray}
s =-\tau_{\rm Q}e^{-t/\tau_{Q}};
\end{eqnarray}
then, the time-dependent Schr\"{o}dinger equation
\begin{eqnarray}
i\partial _{t}|\varphi \lbrack t]\rangle =H_{\sigma}[t]|\varphi
\lbrack t]\rangle ,
\end{eqnarray}
becomes
\begin{eqnarray}
i\partial _{s}|\varphi (s )\rangle =h(s )|\varphi (s )\rangle ,
\end{eqnarray}
with $|\varphi (s )\rangle =|\varphi \lbrack t(s )]\rangle $ and $%
h(s )=h[t(s )]$. Consequently, the time interval of the dynamics
$t\in \lbrack
0,\infty ]$ is mapped into $s \in \lbrack -\tau_{\rm Q%
},0].$

As stated above, the $z$ component of ${\vec{B}}({\vec{r}},t)$
takes large positive and negative values, respectively, at $t=0$
and $t=\infty $. Therefore, at $s =-\tau_{\rm Q}$ and $s =0$, the
condition
\begin{eqnarray}
\left\vert g^{\rm Q}_{ z }-e^{-\xi t(s)}g^{\rm A%
}_{z }\right\vert \gg g_{\perp },
\end{eqnarray}
is satisfied while $g^{\rm Q}_{ z }-e^{-\xi t(s)}g^{\rm A%
}_{z }$ takes positive and negative values, respectively. In the
Landau-Zener approximation, a linear approximation is always
assumed for the different energy levels. We find the value $s
_{0}$ at the crossing point $t_{0}$ is given by
\begin{eqnarray}
s_{0}=-\tau_{\rm Q}q(t_{0})=-\tau_{\rm Q}\left( \frac{g^{%
\rm A}_{z}}{g^{\rm Q}_{ z }}\right) ^{%
\frac{1}{\xi \tau_{\rm Q}}}.
\end{eqnarray}
 At $s =s_{0}$ when the longitudinal B-field vanishes
\begin{eqnarray}
\left( g^{\rm Q}_{ z }-e^{-\xi t(s_{0})}g^{%
\rm A}_{z }\right) =0,
\end{eqnarray}
a linear approximation to the energy levels simply leads to%
\begin{eqnarray}
\left( g^{\rm Q}_{ z }-e^{-\xi t(s)}g^{\mathrm{%
A}}_{z }\right) \approx v(s-s _{0}),
\end{eqnarray}
with
\begin{eqnarray*}
v
&=&-g^{\rm A}_{ z }\tau_{\rm Q}(\xi \tau_{\rm Q%
})\left( \frac{g^{\rm A}_{ z }}{g^{\rm Q}_{ z }}\right)
^{-1-\frac{1}{\xi \tau_{\rm Q}}}.
\end{eqnarray*}%
Using the Landau-Zener formula, we immediately find
\begin{eqnarray}
A({\vec{r}})=\exp \left( -\pi \frac{|g_{\perp }|^{2}}{|v|}\right)
.
\end{eqnarray}
For $\tau_{\rm A}\gg \tau_{\rm Q}$, we find $\xi\tau_{\mathrm{Q%
}}\approx 1$, which then leads to
\begin{eqnarray}
A({\vec{r}},t_{\mathrm{int}}) &\simeq &\exp \left( -\pi
\frac{|g_{\perp
}|^{2}g^{\rm A}_{ z }\tau_{\rm Q}}{g_{\rm Q}}%
\right) \nonumber\\
&=&\exp \left( -\pi g_{F}\mu_{\mathrm B}B^{\rm A}_{ z}({\vec{r}})\tau_{%
\rm Q}\frac{B^{\rm Q2}_{ \perp  }\left( {\vec{r}}%
\right) }{2B^{\rm Q2}_{z}\left( {\vec{r}}\right) }e^{-%
\frac{t_{\mathrm{int}}}{\tau_{\rm A}}}\right). \hskip 24pt
\label{estimation1}
\end{eqnarray}%
Obviously for a large enough time interval $t_{\mathrm{int}}$ such that $%
e^{-t_{\mathrm{int}}/\tau_{\rm A}}\ll 1$, we have $A({\vec{r}},t_{%
\mathrm{int}})\approx 1$ and $P_{2}({\vec{r}})\approx 1$. A single component
condensate remains a single component one. In fact, if $e^{-t_{\mathrm{int}%
}/\tau_{\rm A}}\ll 1$, the bias field has already decreased to
zero when the QUIC field is switched off. Thus, during the
switching-off of the QUIC field, the direction of the B-field does
not change and nonadiabatic transitions cannot occur.

\begin{figure}[tbp]
\includegraphics[width=3.25in]{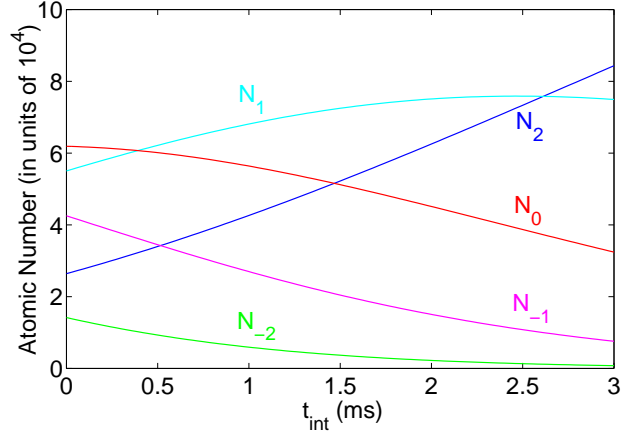}\vspace{0.3cm}
\caption{(Color online)
A typical dependence of the population distributions on the time $%
t_{\mathrm{int}}$. The unit of the atomic number is $10^{-4}$.}
\label{fig2}
\end{figure}

In the experiment of PKU, \cite{chenshuai}, the various trap
parameters take the
following values: $B^{\rm Q}_{z}(0)=9$ (Gauss), $B^{%
\rm A}_{z}=7.45$ (Gauss),
$B^{\rm Q\prime\prime}_{z}=4.9\times 10^{2}$ (Gauss-cm$^{-2}$), $B^{\mathrm{%
Q}\prime}_{\perp}=3.0\times 10^{2}$ (Gauss-cm$^{-1}$%
), $\tau_{\rm Q}=40$ (${\mu }$s), $\tau_{\rm A}=3$ (ms). Before
switching-off, the center of the QUIC trap is at
${\vec{r}}_{0}=(0,5\mathrm{\mu m},0)$. Substituting the above
coefficient $A({\vec{r}})$ of Eq. (\ref{estimation1}) into Eq.
(\ref{probability}), we arrive at a simple estimate for the
population distribution $P_{M_{F}}({\vec{r}}_{0})$. More precisely,
we can estimate the population distribution with the
following,
\begin{eqnarray}
N_{M_{F}}=\int P_{M_{F}}({\vec{r}})\rho ({\vec{r}})d{\vec{r}},
\end{eqnarray}
where $\rho ({\vec{r}})$ is the density profile of the trapped gas
cloud. Figure \ref{fig2} shows the typical dependence of such
result on the time interval $t_{\mathrm{int}}$. In the calculation
of Figure \ref{fig2}, we set $\rho ({\vec{r}})$ to be the atomic
density distribution given by the Thomas-Fermi approximation
corresponding to the initial values of the QUIC field and the bias
field. Namely, the spatial motion of the atoms is omitted. This
approximation is based on the fact that the decay time
$(\rm{3ms})$ is shorter than the period ($\sim$4.5-7.5ms) of the
trap potential and seems to be a bit crude. To obtain a more
accurate estimation of the atomic population, variations of the
atomic spatial distribution in the decay process of the bias field
should be considered fully .

The above result is based on the approximation that atoms do not
move during the switching-off process. If atomic motion is
included, more accurate population distribution can be calculated
by solving the multiple-component Gross-Pitaevskii equation
including the time-dependent B-field. A detailed comparison of
these two approaches is given in Ref. \cite{peng}. Overall, we
find the approximate Landau-Zener solution discussed here holds
well for the parameter regimes of the experiment \cite{chenshuai}.

\section{An alternative scenario}

The QUIC coil consists of a pair of quadruple coils and an Ioffe
coil. The B-field ${\vec B}^{\mathrm Q}$ of the QUIC trap is the
sum of the B-fields ${\vec{B}}^{\mathrm{qd}}$ from the quadruple
coils and ${\vec{B}}^{\mathrm{If}}$ from the Ioffe coil. In the
previous section, we simply assumed the B-fields generated by
these two sets of coils decrease synchronously after shutting off
electric currents . However, as was discovered in the experiment
\cite{chenshuai}, the magnetic fields ${\vec{B}}^{\mathrm{qd}}$
and ${\vec{B}}^{\mathrm{If}}$ do not always decay with the same
time constant despite the fact that the two sets of coils forming
the QUIC trap are in series.
Assuming different time constants for the decay of ${\vec{B}}^{%
\mathrm{qd}}$ and ${\vec{B}}^{\mathrm{If}}$, the finial population
distribution needs to be re-calculated.

Before the QUIC field is switched off,
the components of the B-fields ${\vec{B}}^{\mathrm{qd}}$ and ${\vec{B}}^{%
\mathrm{If}}$ are functions of the atomic position, explicitly given by
\begin{equation}
\begin{array}{lll}
B^{\mathrm{qd}}_{z}({\vec r})&=&B^{\mathrm{qd}\prime}(z-z_{0}),  \nonumber\\
B^{\mathrm{If}}_{z}({\vec r})&=&B^{\rm Q\prime\prime}_{z }z^{2}
-B^{\mathrm{qd}\prime}(z-z_{0})+B^{\rm Q}_{z}(0),   \\
B^{\mathrm{qd}}_{x}({\vec r})&=&-2B^{\mathrm{qd}\prime}x,  \\
B^{\mathrm{If}}_{x}({\vec r})&=&(-B^{\mathrm{Q\prime}}_{\perp
}+2B^{\mathrm{qd}\prime%
})x, \\
B^{\mathrm{qd}}_{y}({\vec r})&=&B^{\mathrm{qd}\prime}y,  \\
B^{\mathrm{If}}_{y}({\vec r})&=&(B^{\mathrm{Q\prime}}_{\perp
}-B^{\mathrm{qd}\prime })y,
\end{array}
\end{equation}
where $z_{0}$ is the distance between the center of the QUIC trap
(in the absence of gravity) and the center of the quadruple trap.

In their experiment \cite{chenshuai}, $B^{\mathrm{qd}\prime}$ is
about $150$ (Gauss-cm$^{-1}$) and $z_{0}$ is $0.75$ (cm).
Therefore, in the region near the center of the QUIC trap,
we have $B^{\mathrm{qd}}_{z}=-107$ (Gauss). From $B^{\rm Q%
}_{z}(0)=9$ (Gauss), we find $B^{\mathrm{If}}_{z}=116$ (Gauss).

If ${\vec{B}}^{\mathrm{If}}$ and ${\vec{B}}^{\mathrm{qd}}$
decrease with different time constants $\tau_{\mathrm{If}}$ and
$\tau_{\mathrm{qd}}$ after switching-off, the B-field from the
quadruple coils becomes
\begin{eqnarray}
(B^{\mathrm{qd}}_{z}%
\mathbf{e}_{z}+B^{\mathrm{qd}}_{y}\mathbf{e}_{y}+B^{\mathrm{qd}}_{x}\mathbf{e}_{x}) e^{-t/\tau_{%
\mathrm{qd}}}
\end{eqnarray}
and the B-field created by the Ioffe coil becomes
\begin{eqnarray}
(B^{\mathrm{If}}_{z}%
\mathbf{e}_{z}+B^{\mathrm{If}}_{y}\mathbf{e}_{y}+B^{\mathrm{If}}_{x}\mathbf{e}_{x}) e^{-t/\tau_{%
\mathrm{If}}}.
\end{eqnarray}

In this section, we assume the time interval $t_{\mathrm {int}}$
between the switching-off of the B-fields ${\vec B}^{Q}$ and
${\vec B}^{A}$ is sufficiently long, i.e., when the QUIC field is
switched off, the bias field has already decreased to zero. Since
the B-fields $B^{\mathrm {qd}}_{z}$ and $B^{\mathrm{If}}_{z}$ have
different signs, at a time $ \tilde {t}_{0}$ the condition
\begin{eqnarray}
B^{\mathrm{If}}_{z}e^{-\tilde {t}_{0}/\tau_{\mathrm{If}}}+B^{\mathrm{qd}%
}_{z}e^{-\tilde {t}_{0}/\tau_{\mathrm{qd}}}=0
\end{eqnarray}
can be satisfied and the $z$ component of the total B-field
becomes zero. As
before, nonadiabatic transitions  happen mainly in the temporal region near $%
\tilde {t}_{0}$. Assuming
\begin{eqnarray}
\tau_{\mathrm{If}}=\frac{\tau_{\mathrm{qd}}}{2},
\end{eqnarray}
the nonadiabatic transition probability $\tilde{A}$ in the
spin-$1/2$ case can again be calculated
with the Landau-Zener method used previously provided that $\tilde {t}_{0}<\tau_{\mathrm{if%
}}$. In the present case, we find $\tilde {t}_{0}<0.6\tau_{\mathrm{If}}$.
Thus, we obtain
\begin{eqnarray}
\tilde {A}({\vec r})=\exp \left[ -\pi \frac{g_{F}\mu_{\mathrm B}\,\tau_{\mathrm{If}}\sum_{l}(B^{\mathrm{qd}%
}_{l}|B^{\mathrm{If}}_{z}|+B^{\mathrm{If}}_{l}|B^{\mathrm{qd}}_{z}|)^{2}}{|B^{\mathrm{If}}_{z}|^3}\right],
\hskip4pt \label{newA}
\end{eqnarray}
where $l=x,y$, the counterpart of the parameter $A$ in the
first scenario.

Now we discuss a special case. We assume the bias field is
switched off adiabatically, such that the atomic cloud follows the
variation of the total B-field, and moves to the region near the
center ${\vec r}_1$ of the QUIC trap (in the absence of the bias
field). Since the Landau-Zener method is based on the assumption
that the atoms are located in the region where the QUIC field lies
approximately along the $z$ axis before being switched off, the
factor $\tilde {A}$ given in Eq. (\ref{newA}) is applicable if the
trap center ${\vec r}_1$ is near the $z$ axis so that $|B^{\mathrm
{Q}}_{(x/y)}|$ is much smaller than $|B^{\mathrm Q}_{z}|$. For
practical values of ${\vec B}^{\mathrm{If}}$ and ${\vec
B}^{\mathrm{qd}}$, the above condition is satisfied and a good
estimate for the transition probability can again be given by Eq.
(\ref{newA}).

From the directions of the electric currents in the quadruple
coils and the Ioffe coil as shown in Figure \ref{fig1}, the
B-fields $B^{\mathrm{If}}_{y}$ and $B^{\mathrm {qd}}_{y}$ are
found to have the same sign while the fields $B^{\mathrm{If}}_{z}$
and $B^{\mathrm {qd}}_{z}$ have opposite signs. Therefore, after
switching off, $B^{\mathrm Q}_{y}$ decays much slower than
$B^{\mathrm Q}_{z}$. At time $\tilde {t}_0$ when $B^{\mathrm
Q}_{z}=0$, $B^{\mathrm Q}_{y}$ has the same order of magnitude as
its initial value. Therefore, if in the region near ${\vec r}_1$,
$B^{\mathrm Q}_{y}$ is sufficiently large, the direction of ${\vec
B}^{Q}$  may be changed very slowly during the switching-off
process of ${\vec B}^{Q}$ so that the atomic spin state can be
adiabatically flipped. Then, we have $\tilde A\approx 0$ and
$P_{-2}\approx 1$. For instance, in the experiment of Ref.
\cite{chenshuai}, ${\vec r}_1=(0,30\mu{\mathrm m},0)$ and
$B^{\mathrm Q}_{y}\sim 0.45 $(Gauss). In this case $85$ percent of
the atoms can be switched to the state $|M_F=-2\rangle$.

A second case of some interest is when the bias field is switched off suddenly.
Once the bias field is turned off, the atoms begin to oscillate in the
new QUIC trap centered at ${\vec r}_{1}$. If the time interval
between the switching off of ${\vec B}^{\mathrm A}$ and ${\vec
B}^{\mathrm Q}$ is $\tilde t_{\mathrm {int}}$, then the population
distribution can be estimated as
\begin{eqnarray}
N_{M_{F}}=\int d{\vec r}\, {\tilde \rho}({\vec r},t_{\mathrm
{int}})P_{M_{F}}[\tilde A({\vec r})].
\end{eqnarray}
Here, ${\tilde \rho}({\vec r},t_{\mathrm {int}})$ is the density
distribution of atoms in the QUIC trap at the time
when the QUIC field is switched off.

\section{Conclusions}

In conclusion, we have presented a detailed theoretical treatment
for the nonadiabatic level crossing dynamics of an atomic spin
coupled to a time dependent magnetic field. When applied to the
condensate experiments in a modified QUIC trap \cite{chenshuai},
our theory provides a satisfactory explanation for the observed
multi-component spinor condensates when the trapping B-fields were
shut-off. In the broad context of condensate wave function
engineering and atom optics with degenerate quantum gases, our
work provides useful insights for experiments. For example, in
some proposals \cite{Machida} and experiments \cite{Ketterle} on
the creation of vortex states in a condensate, the internal atomic hyperfine
state is slated to adiabatically follow the external magnetic field and be changed
from $|m_{F}\rangle_{z}$ to $|-m_{F}\rangle_{z}$. Our method can
then also be used to estimate the nonadiabatic effects in these proposals
and experiments \cite{Machida,Ketterle}.

\vskip 12pt We thank the Peking University atomic quantum gas
group, especially its leader Prof. X. Z. Chen for enlightening
discussions. We thank Prof. Chandra Raman for several helpful
communications. Part of this work was completed while one of us
(L.Y.) was a visitor at the Institute of Theoretical Physics of
the Chinese Academy of Sciences in Beijing, he acknowledges warm
hospitality extended to him by his friends at the Institute. This
work is supported by NSF, NASA, and the Ministry of Education of
China.

\end{document}